\documentstyle[emulateapj]{article}

\newcommand\nstar{$N_*$}
\newcommand\Ha{H$\alpha$}

\newcommand\hii{\ion{H}{2}}
\newcommand\hi{\ion{H}{1}}
\newcommand\etal{et\thinspace al.~}

\newcommand\msol{\rm\,M_\odot}

\newcommand\kms{{\rm\,km\,s^{-1}}}

\def\spose#1{\hbox to 0pt{#1\hss}}
\def\simpropto{\mathrel{\spose{\raise 3pt\hbox{$\propto$}}
     \lower 3.0pt\hbox{$\sim$}}}

\received{17 October 2003}
\accepted{}
\slugcomment{Accepted to the {\bf ASTRONOMICAL JOURNAL} 26 November 2003}

\lefthead{Oey, King, and Parker}
\righthead{Field and cluster OB stars }

\begin{document}

% \title{ {\rm\small *****
% DRAFT version \today\ --- please do not circulate
% SUBMITTED version 21-November-2003 --- please do not circulate
%  ***** } \\ \bigskip
\title{	
Massive Field Stars and the Stellar Clustering Law}

\author{M. S. Oey and N. L. King}
\affil{Lowell Observatory, 1400 W. Mars Hill Rd., 
	Flagstaff, AZ   86001; Sally.Oey@lowell.edu}
% \altaffiltext{1}{}
\author{\medskip and \\ J. Wm. Parker}
\affil{Department of Space Studies, Southwest Research Institute,
	Suite 426, 1050 Walnut St., Boulder, CO\ \ \ 80302}

\begin{abstract}
The distribution of \nstar, the number of OB stars per association or
cluster, appears to follow a universal power-law form $N_*^{-2}$ in
the local Universe.  We evaluate the distribution of \nstar\ in the Small
Magellanic Cloud using recent broadband optical and space-ultraviolet
data, with special attention to the lowest values of \nstar.  We find
that the power-law distribution in \nstar\ continues smoothly down to
\nstar$=1$.  This strongly suggests that the formation of field
massive stars is a continuous process with those in associations, and
that the field stars do not originate from a different star formation
mode.  Our results are consistent with the model that field massive
stars represent the most massive members in groups of smaller stars,
as expected if the clustering law applies to much lower masses
as is expected from the stellar initial mass function (IMF).  These
results are consistent with the {\it simultaneous} existence of a
universal IMF and a universal clustering law.  Jointly, these laws 
imply that the fraction of field OB stars typically ranges from about
35\% to 7\% for most astrophysical situations, with an inverse
logarithmic dependence on the most populous cluster, and hence, on
galaxy size and/or star formation rate.  There are important
consequences for global feedback effects in galaxies:  field stars
should therefore contribute proportionately to the volume of
the warm ionized medium, and equal relative contributions by
superbubbles of all sizes to the interstellar porosity are expected.
\end{abstract}


\section{Introduction}

It is commonly held that most massive, OB stars are found in
stellar clusters, or associations, since their short ($\sim 10$ Myr)
lifetimes are not long enough to permit spatial dispersion from their
natal companions.  However, apparently-isolated, massive field stars
are well-known to exist, including a class of runaway OB stars with
unusually large ($\gtrsim 30\ \kms$) velocities.  While runaway OB
stars are generally believed to be kinematically ejected from a parent
OB association, the ordinary non-runaway field OB stars have been
suggested to originate in a different mode of star formation from
their counterparts in associations.  Suggested differences between
clusters and the field in the stellar initial mass function (IMF)  for
high-mass stars (Massey 2002; Kroupa \& Weidner 2003) support this
possibility.  This contribution explores the relationship between OB
associations and massive field stars.
% (e.g., Li, Klessen, \& Mac Low 2003).

In recent years, it has emerged that the
number of stars  \nstar\ per cluster appears to follow a universal power
law distribution:
\begin{equation}\label{eq_n*}
N(N_*)\ dN_* \propto N_*^{-2}\ dN_*	\quad .
\end{equation}
This has been found empirically for young, massive clusters (e.g.,
Hunter et al. 2003; Zhang \& Fall 1999), super star clusters (Meurer
et al. 1995), globular clusters (e.g., Harris \& Pudritz 1994) and
\hii\ regions (Oey \& Clarke 1998).  The apparent 
universality of this relation is emerging as fundamental (e.g., Oey \&
Mu\~noz-Tu\~non 2003; Oey \& Clarke 1998; Elmegreen \& Efremov 1997),
similar to the constant power-law  relation for the IMF.
If individual, field OB stars have a
fundamentally different origin from clustered OB stars, then this is
likely to be manifest in the distribution of \nstar, near $N_*=1$,
where \nstar\ specifically counts massive stars only (here, $m\gtrsim 10\
\msol$).  In what follows, we examine the form of $N(N_*)$, for small 
\nstar, a regime that has not been investigated to date, to illuminate
the relationship between massive star clustering and massive field stars.

\section{The SMC sample of OB stars}

\begin{figure*}
\vspace*{-0.1truein}
\epsscale{2.0}
%% \plotone{findchart.ps}
\plotone{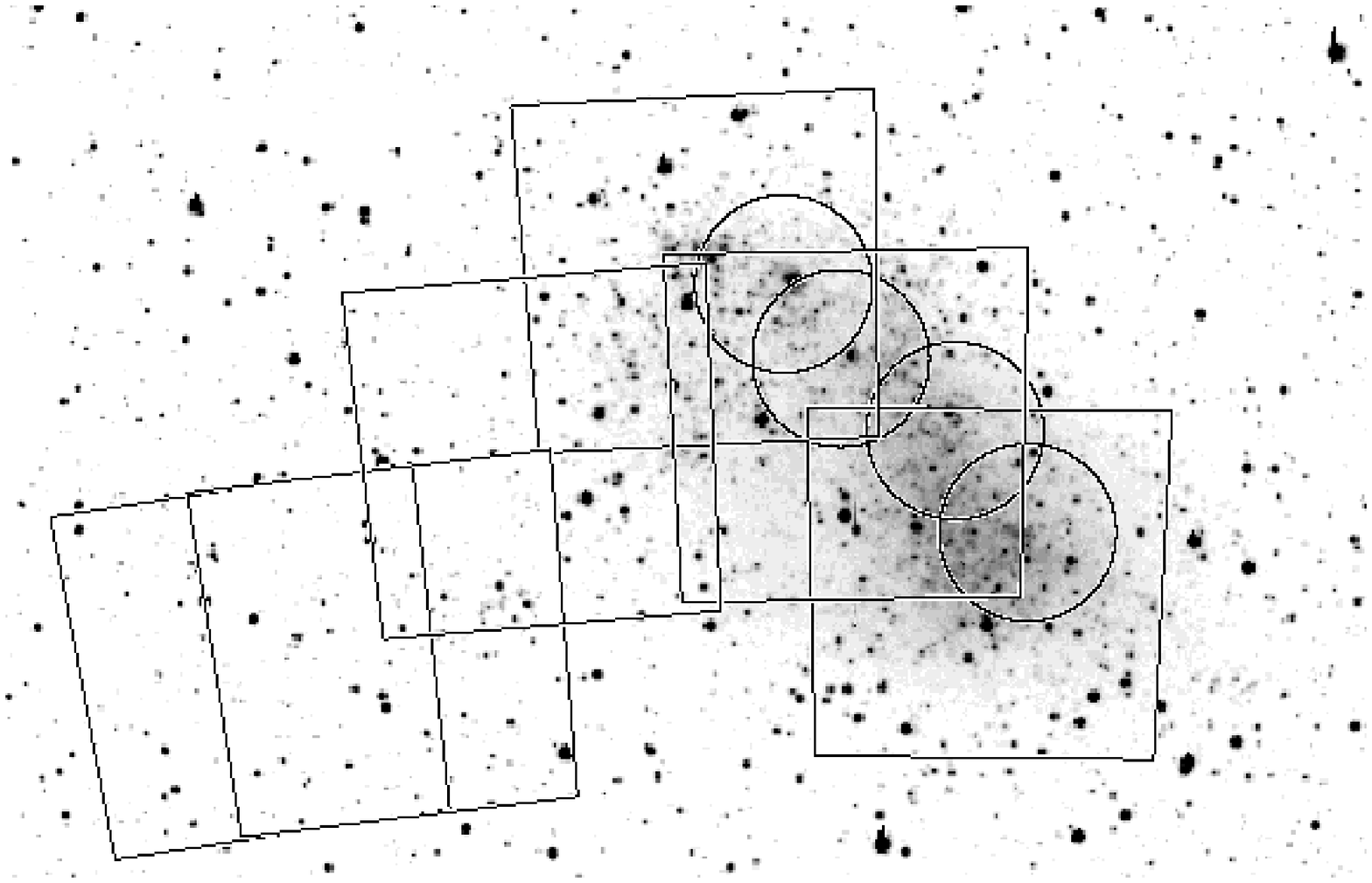}
\caption{Grayscale $R$-band image of the SMC (Bothun \& Thompson 1988)
showing the area coverage for the $UBVR$ survey (Massey et al. 2002;
rectangular regions) and the {\sl UIT} survey (Parker et al. 1998;
circular regions).  The optical fields are $79\arcmin$ square, and the
{\it UIT} fields are $37\arcmin$ in diameter.  North is up, east to
the left.
\label{f_areas}}
\end{figure*}

A study of OB star clustering properties and field stars requires high
spatial resolution and essentially complete detection of the OB stars
over a large area of a given galaxy.  The Magellanic Clouds are
optimal target galaxies by virtue of their proximity and high Galactic
latitude.  Indeed, some of the earliest work on OB associations and
massive star censuses was carried out for the Magellanic Clouds by
Feast et al. (1960), Sanduleak (1969), Lucke \& Hodge (1970),
Azzopardi \& Vigneau (1982), and Hodge (1985).  More recently, the
$UBVR$ survey data of Massey (2002) for the Small 
Magellanic Cloud (SMC), together with UV photometry (Parker et al.
1998) from the {\sl Ultraviolet Imaging Telescope (UIT)},
provide  unprecedented depth, resolution, completeness, and broadband
coverage for the massive star population over a large fraction of the
SMC.  The area observed by {\sl UIT} is smaller than the optical
survey, but covers the SMC bar, which includes most of the active star
formation in that galaxy.  Figure~\ref{f_areas} shows the regions
covered by the two surveys.  SMC OB associations have previously been
identified and catalogued by Hodge (1985) and Battinelli (1991), the
latter from the photographic OB census of Azzopardi \& Vigneau
(1982).  Here, we re-examine the clustering properties of these
massive stars in the SMC, using the modern datasets to first
systematically identify OB stars, and then to systematically identify
groups or associations of them.  We will especially focus on the
statistical properties of the low-\nstar\ regime. 

\cite{par98} provide {\sl UIT} photometry in the B5 filter
($\lambda_{\rm eff}$ = 1615 \AA, $\Delta\lambda$ = 225 \AA), 
which recently has been recalibrated by Parker, Cornett, \& Stecher
(2004, in preparation).  The
$UBVR$ photometry of \cite{mas02} was observed through a Harris filter
set and transformed to the Johnson-Cousins system (Landolt 1992).    
Reddening-free $Q$ indices were computed using the extinction law 
of \cite{ccm89} with a total to selective extinction ratio $R_V$ =
3.1.  Different combinations of filters were used to best select two
different stellar samples:  Stars with initial masses $\gtrsim 10\
\msol$, corresponding to spectral types of B0 V, B0.5 I and earlier,
were selected on the basis of their optical photometry only;
and stars with initial masses $\gtrsim 20\ \msol$, corresponding to
spectral types of O9 V, B0 I and earlier were selected on the basis of
both their $UBVR$ and {\sl UIT} photometry.  We refer to the former 
as the OB sample, and the latter as the O-star sample, although we
emphasize that neither sample consists of spectroscopically confirmed stars. 
Our criteria are as follows, using an SMC distance modulus DM$=18.88$
(Dolphin et al. 2001) and global average extinction $A_B = 0.53$:
\begin{eqnarray}\label{eq_Ocritfirst}
m_{B} & \leq & -4.2 + {\rm DM} + A_B \nonumber \\
      & \leq & 15.21 \\
Q_{U-R,B-R} & \leq & -0.84, \label{eq_Ocritsecond}\\
Q_{B5-U,B-R} & \leq & -1.45, \\
Q_{B5-V,U-V} & \leq & 0.03, 
\label{eq_Ocritlast}
\end{eqnarray}

where 
\begin{eqnarray}
Q_{U-R,B-R} & = & (m_U - m_R) - \frac{A_U - A_R}{A_B - A_R} 
	(m_B - m_R) \nonumber \\
            & = & (m_U - m_R) - 1.396\ (m_B - m_R)\ , \\
Q_{B5-U,B-R} & = & (m_{B5} - m_U) - \frac{A_{B5} - A_U}
	{A_B - A_R} (m_B - m_R) \nonumber \\
            & = & (m_{B5} - m_U) - 1.668\ (m_B - m_R)\ , \\
Q_{B5-V,U-V} & = & (m_{B5} - m_V) - \frac{A_{B5} - A_V}
	{A_U - A_V} (m_U - m_V) \nonumber \\
            & = & (m_{B5} - m_V) - 2.718\ (m_U - m_V)\ . 
\end{eqnarray}
The OB sample is selected from equations~\ref{eq_Ocritfirst} and
\ref{eq_Ocritsecond} only, while the O-star sub-sample meets all of
equations~\ref{eq_Ocritfirst} to \ref{eq_Ocritlast}.

Table~\ref{tab_ostarlist} shows all the stars having both $UIT$ B5
photometry, $UBVR$ photometry, and spectral classifications listed by
Massey (2002), excluding
stars with objective prism classifications and those identified as
uncertain spectral types.  The ID number and classification from
Massey (2002) are shown in Columns 1 and 2.  Column 3 shows whether
the star met the O-star selection criteria above. 
Our attempts to select only stars earlier than O9 V and B0 I were
successful for 27 out of 31 stars; of the 4 not selected, one had
nebular emission affecting the photometry and two others had weak
metal lines (designation W in spectral type).  The false positives
were 8 out 26 stars; the majority of the false positives were 
weak lined stars (W), emission line stars, and peculiar stars.
%initial masses >= 20 Msun ---> earlier than O9V and B0I
%initial masses >= 10 Msun ---> earlier than B0V and B0.5I

\begin{deluxetable}{clc|clc}
\footnotesize
\tablewidth{0pt}
\tablecolumns{6} 
\tablecaption{Comparison of O-star criteria\label{tab_ostarlist}} 
\tablehead{
\colhead{Cat ID\tablenotemark{a}} & {Sp Type\tablenotemark{b}} & {Identified\tablenotemark{c}} &
\colhead{Cat ID\tablenotemark{a}} & {Sp Type\tablenotemark{b}} & {Identified\tablenotemark{c}} \\
\colhead{ } & \colhead{ } & \colhead{(y or n)} &
\colhead{ } & \colhead{ } & \colhead{(y or n)} }
\startdata
 16828 & O5 III(f) & y  & 53382 & B0 IIWW   & y \\
 38024 & O5 V      & y  & 54456 & B0 IW     & y \\
  5869 & O7 V      & y  & 54958 & B0 IIIWW  & y \\
 16056 & O7 III    & y  &  9079 & B1.5 V    & n \\
 16885 & O7 III    & y  &  9488 & B1 I      & n \\
 17457 & O7 III    & y  & 18700 & B1 Ib     & y \\
 17927 & O7.5Iaf+  & y  & 37732 & B1 IIW    & n \\
 19650 & O7.5 III  & y  & 42740 & B1 III    & n \\
 22837 & O7.5 V    & y  & 43807 & B1 V      & n \\
 40341 & O7 III    & y  & 44784 & B1 III    & n \\
 43724 & O7 V      & y  & 45114 & B1 III    & y \\
 45521 & O7 If     & y  & 45438 & B1 Ve     & n \\
 46035 & O7 III    & y  & 45809 & B1 III    & n \\
 52170 & O7 V      & y  & 50609 & B1 III    & n \\
 27731 & O8.5 V    & y  & 51009 & B1.5 III  & n \\
 43197 & O8.5 V    & y  & 53225 & B1 V      & n \\
 43734 & O8 V      & n  & 26901 & B2 IIIW   & y \\
 49580 & O8.5 V    & y  & 32907 & B2 II     & n \\
 50825 & O8 Vn     & y  & 50826 & B2.5 III  & n \\
 53373 & O8.5 V    & y  & 54818 & B2 I      & n \\
  6406 & O9+NEB    & n  & 24929 & B3 I      & n \\
 11777 & O9 II     & y  & 43844 & B3 I      & n \\
 40610 & O9 V      & y  & 44828 & B3 I      & n \\
 45677 & O9 III    & y  & 47028 & B3 I      & n \\
 47540 & O9 III    & y  & 49606 & B3 I      & n \\
  3459 & B0 IWW    & n  & 45722 & B8 Ve     & y \\
 13075 & B0.5 III  & y  & 50475 & B8 Iab    & n \\
 13831 & B0.5 V    & y  & 53084 & B8-A0 I   & n \\
 15503 & B0.5 V    & n  &  5391 & B pec     & y \\
 15742 & B0 IIWW   & y  & 21801 & B pec     & y \\
 18614 & B0 IIW    & n  & 24914 & B extr    & y \\
 20656 & B0 I      & y  & 40851 & B extr    & y \\
 23702 & B0 IW     & y  & 10915 & A0 Ia     & n \\
 41648 & B0.5 III  & n  & 43215 & A0 I      & n \\
 43686 & B0.5 V    & n  & 48732 & A0 I      & n \\
 43758 & B0.2 V    & n  & 52992 & A1 I      & n \\
 51575 & B0 III    & y  & \nodata & \nodata & \nodata \\
\enddata
\tablenotetext{a}{Catalog number from \cite{mas02}}
\tablenotetext{b}{Spectroscopic type compiled by \cite{mas02}}
\tablenotetext{c}{Identified by O-star criteria
equations~\ref{eq_Ocritfirst} -- \ref{eq_Ocritlast}}
\end{deluxetable}

\section{Identification of OB associations}

\begin{figure*}
\vspace*{-0.1in}
\epsscale{2.0}
%\plotone{d3fig.ps}
\plotone{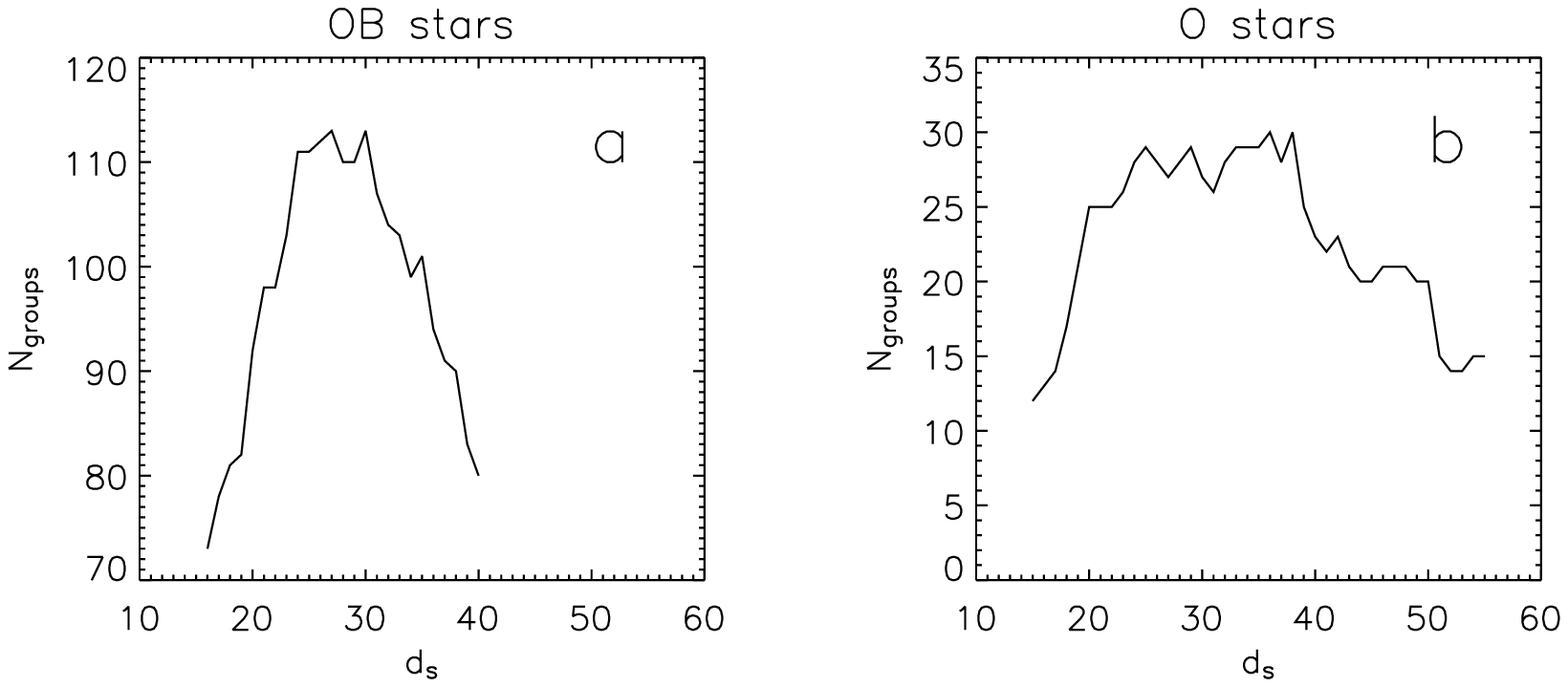}
%\vspace*{-6.2in}
\vspace*{-5.5in}
\caption{Number of groups with $N_*\geq3$ as a function of clustering
distance $d_s$ for the OB sample (panel $a$) and O-star sample (panel
$b$). 
\label{f_ds} }
\end{figure*}

Having identified two samples of 1360 OB and 382 O stars,
we then used the friends-of-friends algorithm described by Battinelli
(1991) to identify the associations of these stars.  Our parameter
\nstar\ refers specifically to counts of OB stars from these samples,
identified as described above.  Battinelli's algorithm adopts as the
clustering distance $d_s$ between associated members the value of
$d_s$ that maximizes the number of clusters for $N_* \geq 3$.  All
stars within $d_s$ of another member star are defined to be within the
same group.  (We do not distinguish between ``group,''
``association,'' and ``cluster'' in this work.)  Figure~\ref{f_ds}
shows the number of clusters $N$ as a function $d_s$ near the
peak in $N$ for the OB sample (Figure~\ref{f_ds}$a$) and O-star
sample (Figure~\ref{f_ds}$b$).  For the OB sample, the characteristic
$d_s=28$ pc ($97\arcsec$) is straightforward to determine, but for the
O-star sample, it is apparent that $N$ is less sensitive to
$d_s$.  We adopt $d_s = 34$ pc ($117\arcsec$) for the O-star sample,
corresponding to the mean value over the considered range;
the larger value of $d_s$ as compared to the OB sample is consistent
with the lower stellar density 
for O stars.  Although the clustering properties are relatively
insensitive to the choice of $d_s$ within a few pc, we caution that
larger values of $d_s$ will result in flatter slopes for the
clustering law $N(N_*)$ (e.g., equation~\ref{eq_n*}), as discussed
below in \S 4.  Our values of $d_s$ are much smaller than Battinelli's
value of 60 pc; this is consistent with a much greater completeness in
our data, yielding a higher density of OB stars, and therefore smaller
clustering distance. 

\begin{deluxetable}{rcccrr}
\footnotesize
% \tiny
\tablewidth{0pt}
\tablecolumns{6} 
\tablecaption{Groups from the candidate OB sample \label{t_OBgrpctrs}} 
\tablehead{
\colhead{OB grp} & {RA} & {Dec} &
\colhead{$D$(min)} & {$D$(pc)} & \nstar }
\startdata
% #group   racen    deccen     diam(min) diam(pc) nstars
  20 & 11.3029 & --73.3820 &  1.38 &   23.93 &   4 \\
  24 & 11.3639 & --73.2441 &  3.39 &   58.90 &   5 \\
  29 & 11.5932 & --73.1044 &  1.45 &   25.21 &   4 \\
  30 & 11.5857 & --73.4439 &  1.69 &   29.37 &   3 \\
  31 & 11.5817 & --73.2120 &  0.38 &    6.64 &   3 \\
  38 & 12.0545 & --73.1842 & 12.91 &  224.68 &  70 \\
  40 & 11.7724 & --73.2863 &  0.81 &   14.05 &   3 \\
  41 & 11.8078 & --73.3658 &  1.97 &   34.36 &   4 \\
  45 & 12.0111 & --73.3695 &  1.92 &   33.33 &   4 \\
  47 & 12.2249 & --73.4337 &  6.93 &  120.56 &  26 \\
  64 & 12.2657 & --72.0873 &  1.75 &   30.42 &   4 \\
  73 & 12.4821 & --73.3470 &  3.95 &   68.70 &  16 \\
  74 & 12.4058 & --73.0955 &  0.85 &   14.84 &   3 \\
  75 & 12.4251 & --72.8051 &  1.02 &   17.81 &   3 \\
  86 & 12.5262 & --73.2635 &  0.90 &   15.73 &   3 \\
  91 & 12.6156 & --72.8806 &  1.16 &   20.18 &   4 \\
  98 & 12.6898 & --73.4584 &  1.47 &   25.53 &   6 \\
  99 & 12.7466 & --72.6852 &  3.78 &   65.71 &   9 \\
 100 & 12.6862 & --72.7800 &  1.99 &   34.61 &   5 \\
 101 & 12.7490 & --73.2845 &  3.42 &   59.56 &  10 \\
 104 & 12.7620 & --73.1756 &  1.14 &   19.90 &   4 \\
 107 & 13.1672 & --72.7186 & 12.48 &  217.09 &  55 \\
 110 & 12.8384 & --73.3445 &  2.35 &   40.93 &   7 \\
 112 & 12.8514 & --72.9475 &  1.46 &   25.40 &   3 \\
 115 & 12.8844 & --72.6976 &  1.20 &   20.94 &   3 \\
 118 & 12.9530 & --72.8498 &  3.02 &   52.62 &   7 \\
 119 & 12.8984 & --72.4770 &  0.91 &   15.80 &   3 \\
 120 & 12.9375 & --73.2152 &  2.21 &   38.45 &   8 \\
 121 & 12.9762 & --73.3923 &  1.59 &   27.72 &   3 \\
 123 & 13.1031 & --72.1594 &  5.99 &  104.24 &  15 \\
 125 & 13.0386 & --72.6643 &  2.14 &   37.23 &   6 \\
 131 & 13.0357 & --72.3948 &  1.05 &   18.26 &   3 \\
 137 & 13.1042 & --73.3233 &  1.57 &   27.33 &   4 \\
 142 & 13.3029 & --73.1472 &  3.52 &   61.32 &   8 \\
 150 & 13.2429 & --72.9064 &  1.64 &   28.61 &   3 \\
 152 & 13.2756 & --73.1909 &  1.06 &   18.42 &   3 \\
 158 & 13.4218 & --72.5561 &  1.30 &   22.63 &   3 \\
 160 & 13.4898 & --72.2771 &  3.20 &   55.66 &   6 \\
 170 & 13.5753 & --72.3525 &  3.01 &   52.33 &   9 \\
 174 & 13.6399 & --72.4905 &  1.70 &   29.66 &   4 \\
 195 & 13.9908 & --72.3949 &  1.86 &   32.43 &   4 \\
 197 & 13.9913 & --72.2634 &  1.06 &   18.43 &   3 \\
 204 & 14.0291 & --72.4610 &  0.71 &   12.40 &   3 \\
 211 & 14.0923 & --72.0363 &  0.82 &   14.26 &   3 \\
 228 & 14.3369 & --72.6518 &  1.37 &   23.86 &   5 \\
 233 & 14.4076 & --72.3812 &  2.52 &   43.93 &   5 \\
 235 & 14.4236 & --72.2069 &  1.26 &   22.01 &   3 \\
 237 & 14.5052 & --72.5974 &  1.74 &   30.22 &   5 \\
 241 & 14.5732 & --72.6575 &  2.34 &   40.66 &   8 \\
 244 & 14.7403 & --72.1810 &  5.74 &   99.90 &  42 \\
 247 & 14.6652 & --72.7249 &  2.85 &   49.57 &   5 \\
 248 & 14.5854 & --72.3032 &  0.78 &   13.66 &   3 \\
 255 & 14.6828 & --72.2829 &  0.98 &   17.03 &   3 \\
 256 & 14.7037 & --72.3284 &  1.59 &   27.67 &   3 \\
 263 & 14.8208 & --72.0852 &  1.10 &   19.18 &   3 \\
 265 & 14.8466 & --72.3857 &  1.06 &   18.47 &   3 \\
 270 & 14.8983 & --72.7153 &  1.40 &   24.44 &   3 \\
 277 & 15.0082 & --72.2228 &  2.09 &   36.34 &   4 \\
 279 & 14.9884 & --72.1773 &  1.54 &   26.76 &   3 \\
 294 & 15.1744 & --72.5081 &  1.87 &   32.59 &   4 \\
\enddata
\end{deluxetable}

\setcounter{table}{1}
\begin{deluxetable}{rcccrr}
\footnotesize
% \tiny
\tablewidth{0pt}
\tablecolumns{6} 
\tablecaption{({\it continued})} 
\tablehead{
\colhead{OB grp} & {RA} & {Dec} &
\colhead{$D$(min)} & {$D$(pc)} & \nstar }
\startdata
 303 & 15.2701 & --72.2317 &  2.02 &   35.18 &   4 \\
 315 & 15.3803 & --72.1378 &  1.38 &   23.95 &   3 \\
 325 & 15.4218 & --71.9461 &  1.28 &   22.21 &   3 \\
 327 & 15.5244 & --72.2091 &  2.15 &   37.40 &   5 \\
 331 & 15.5112 & --72.5928 &  0.88 &   15.30 &   3 \\
 333 & 15.5852 & --72.0239 &  3.47 &   60.41 &   7 \\
 337 & 15.8589 & --72.0899 &  8.45 &  147.10 &  79 \\
 340 & 15.6331 & --71.8665 &  2.41 &   41.86 &   5 \\
 344 & 15.7216 & --71.9261 &  3.20 &   55.77 &   4 \\
 345 & 15.6789 & --72.7675 &  1.92 &   33.33 &   4 \\
 347 & 15.7562 & --72.4252 &  2.35 &   40.82 &   6 \\
 348 & 15.7160 & --72.4867 &  2.07 &   35.98 &   4 \\
 351 & 15.7653 & --72.6240 &  1.86 &   32.41 &   3 \\
 354 & 15.9167 & --71.9469 &  3.32 &   57.77 &   8 \\
 357 & 15.8683 & --72.3877 &  1.43 &   24.85 &   3 \\
 365 & 16.0226 & --72.7355 &  2.00 &   34.80 &   8 \\
 366 & 15.9992 & --72.5378 &  1.54 &   26.85 &   4 \\
 367 & 16.0374 & --72.8077 &  0.89 &   15.50 &   3 \\
 369 & 16.0700 & --72.0117 &  2.11 &   36.70 &   4 \\
 373 & 16.1001 & --72.1584 &  0.70 &   12.26 &   3 \\
 376 & 16.1858 & --72.3718 &  1.15 &   20.02 &   3 \\
 377 & 16.2677 & --71.9959 &  3.51 &   60.99 &  19 \\
 382 & 16.2279 & --72.0984 &  1.30 &   22.70 &   4 \\
 383 & 16.2297 & --72.1892 &  0.95 &   16.61 &   3 \\
 385 & 16.2451 & --72.7920 &  1.36 &   23.67 &   5 \\
 388 & 16.3209 & --72.0536 &  2.35 &   40.90 &   4 \\
 391 & 16.3701 & --72.4773 &  3.66 &   63.75 &   8 \\
 399 & 16.5235 & --72.5441 &  1.08 &   18.87 &   4 \\
 415 & 16.7752 & --72.2671 &  2.34 &   40.67 &   5 \\
 416 & 16.7213 & --72.0574 &  0.89 &   15.42 &   3 \\
 420 & 16.8253 & --72.2122 &  2.59 &   45.08 &   5 \\
 421 & 16.7891 & --72.4370 &  3.31 &   57.54 &   6 \\
 426 & 16.8895 & --72.5793 &  2.48 &   43.11 &   5 \\
 429 & 16.9275 & --72.4516 &  1.23 &   21.39 &   4 \\
 439 & 17.0803 & --71.9925 &  1.43 &   24.93 &   3 \\
 444 & 17.2285 & --72.4238 &  3.28 &   56.99 &  10 \\
 449 & 17.3037 & --73.1992 &  1.29 &   22.40 &   4 \\
 460 & 17.7278 & --72.7223 &  1.18 &   20.45 &   3 \\
 464 & 17.8254 & --72.3788 &  0.79 &   13.72 &   4 \\
 480 & 18.4627 & --73.2896 &  1.96 &   34.18 &   8 \\
 486 & 18.8834 & --73.3285 &  6.18 &  107.50 &  24 \\
 496 & 18.9001 & --73.1967 &  0.32 &    5.62 &   3 \\
 529 & 20.6192 & --72.7708 &  0.96 &   16.66 &   3 \\
 538 & 21.0626 & --73.1555 &  1.95 &   33.88 &   8 \\
 539 & 21.0881 & --73.2422 &  1.88 &   32.75 &   4 \\
 546 & 21.2965 & --73.2821 &  0.90 &   15.68 &   3 \\
 547 & 21.3705 & --73.3945 &  1.73 &   30.09 &   4 \\
 550 & 21.5905 & --73.2984 &  0.96 &   16.65 &   3 \\
 568 & 22.3738 & --73.5585 &  0.52 &    8.98 &   3 \\
 569 & 22.6252 & --73.4156 &  3.88 &   67.59 &  11 \\
 576 & 22.6252 & --73.4156 &  3.88 &   67.59 &  11 \\
\enddata
\end{deluxetable}

\begin{deluxetable}{rcccrr}
\footnotesize
\tablewidth{0pt}
\tablecolumns{6} 
\tablecaption{Groups from the candidate O-star sample \label{t_Ogrpctrs}} 
\tablehead{
\colhead{O grp} & {RA} & {Dec} &
\colhead{$D$(min)} & {$D$(pc)} & \nstar }
\startdata
   8 & 11.9117 & --73.1385 &  4.87 &   84.71 &  11 \\
  11 & 11.8196 & --73.2295 &  1.11 &   19.35 &   3 \\
  13 & 12.0083 & --73.3533 &  2.64 &   45.86 &   4 \\
  19 & 12.2742 & --73.2744 &  3.57 &   62.11 &   6 \\
  24 & 12.2523 & --73.1099 &  1.41 &   24.49 &   3 \\
  26 & 12.4694 & --72.6139 &  3.03 &   52.78 &   4 \\
  38 & 12.6156 & --72.8806 &  1.16 &   20.18 &   4 \\
  41 & 13.0895 & --72.7004 & 13.76 &  239.45 &  57 \\
  45 & 12.9077 & --72.7988 &  5.06 &   88.03 &  12 \\
  50 & 12.9830 & --72.8530 &  2.30 &   40.07 &   4 \\
  51 & 12.9693 & --72.5435 &  1.60 &   27.84 &   4 \\
  62 & 13.4522 & --72.5525 &  2.19 &   38.11 &   4 \\
  67 & 13.5963 & --72.3412 &  5.02 &   87.31 &  12 \\
  68 & 13.6438 & --72.4994 &  2.52 &   43.85 &   5 \\
  76 & 13.8855 & --72.4820 &  1.74 &   30.28 &   3 \\
  78 & 13.9908 & --72.3949 &  1.86 &   32.43 &   4 \\
  81 & 14.2089 & --72.4833 &  5.75 &  100.05 &  11 \\
  84 & 14.0923 & --72.0363 &  0.82 &   14.26 &   3 \\
  97 & 14.4076 & --72.3812 &  2.52 &   43.93 &   5 \\
 102 & 14.5581 & --72.6512 &  1.87 &   32.61 &   4 \\
 104 & 14.7329 & --72.1834 &  5.90 &  102.61 &  26 \\
 107 & 14.6211 & --72.7179 &  2.00 &   34.71 &   3 \\
 109 & 14.6682 & --72.2968 &  2.38 &   41.43 &   5 \\
 121 & 15.0359 & --72.2274 &  1.72 &   29.92 &   3 \\
 124 & 15.0641 & --72.5572 &  1.42 &   24.74 &   3 \\
 130 & 15.2501 & --72.2311 &  2.68 &   46.59 &   5 \\
 143 & 15.5244 & --72.2091 &  2.15 &   37.40 &   5 \\
 145 & 15.5583 & --72.0413 &  1.55 &   26.96 &   3 \\
 146 & 15.7173 & --72.1075 &  5.36 &   93.19 &  17 \\
\enddata
\end{deluxetable}

\begin{deluxetable}{cccccccccccc}
\footnotesize
\tablewidth{0pt}
\tablecolumns{12} 
\tablecaption{Group membership for the candidate OB
sample\tablenotemark{a}
\label{t_4}} 
\tablehead{
\colhead{RA} & {Dec} & $U$ & $U$ err & $B$ & $B$ err & $V$ & $V$ err
& $R$ & $R$ err & ID & \colhead{OB grp} }
\startdata
  10.1171 & --73.5425 & 14.01 &  0.05 & 14.96 &  0.04 & 15.00 &  0.03 &
14.96 &  0.07 &  107  &   1 \\
  10.1832 & --73.4063 & 14.40 &  0.05 & 15.18 &  0.04 & 15.12 &  0.03 &
14.86 &  0.07 &  298  &   2 \\
  10.3880 & --73.4256 & 14.24 &  0.06 & 15.15 &  0.06 & 15.28 &  0.04 &
15.30 &  0.11 & 1037  &   3 \\
  10.5417 & --73.2323 & 13.44 &  0.03 & 14.42 &  0.03 & 14.60 &  0.02 &
14.69 &  0.05 & 1600  &   4 \\
  10.5515 & --73.3866 & 14.26 &  0.05 & 15.19 &  0.04 & 15.15 &  0.03 &
15.04 &  0.08 & 1631  &   5 \\
\enddata
\tablenotetext{a}{The complete version of this table is in the electronic edition of
the Journal.  The printed edition contains only a sample.}
\end{deluxetable}

\begin{deluxetable}{cccccccccccccc}
\footnotesize
\tablewidth{0pt}
\tablecolumns{14} 
\tablecaption{Group membership for the candidate O-star sample\tablenotemark{a}
\label{t_5}} 
\tablehead{
\colhead{RA} & {Dec} & $U$ & $U$ err & $B$ & $B$ err & $V$ & $V$ err
& $R$ & $R$ err & B5 & B5 err & ID & \colhead{O grp} }
\startdata
  11.1440 & --73.1597 & 13.78 & 0.05 & 14.70 & 0.04 & 14.82 & 0.03 &
14.74 & 0.07 & 11.92 & 0.05 & 4424 &  1 \\
  11.2379 & --73.0130 & 13.19 & 0.03 & 14.30 & 0.03 & 14.36 & 0.02 &
14.29 & 0.04 & 11.16 & 0.09 & 4922 &  2 \\
  11.2438 & --73.0213 & 14.05 & 0.05 & 14.94 & 0.04 & 14.91 & 0.03 &
14.86 & 0.07 & 12.19 & 0.05 & 4949 &  2 \\
  11.3259 & --73.2564 & 12.45 & 0.02 & 13.36 & 0.01 & 13.31 & 0.01 &
13.14 & 0.02 & 10.25 & 0.08 & 5391 &  3 \\
  11.3867 & --73.0773 & 13.16 & 0.04 & 14.15 & 0.04 & 14.30 & 0.03 &
14.28 & 0.06 & 10.85 & 0.06 & 5718 &  4 \\
\enddata
\tablenotetext{a}{The complete version of this table is in the electronic edition of
the Journal.  The printed edition contains only a sample.}
\end{deluxetable}

Table~\ref{t_OBgrpctrs} presents the groups having at least 3 stars
that are identified from our OB sample.  Column~1 shows the group ID
number, columns~2 and 3 give the group centroid position in decimal
degrees (J2000.0), columns~4 and 5 give the group diameter $D$ in
arcmin and pc, respectively, and column~6 lists the number of stars in
the group.  Table~\ref{t_Ogrpctrs} presents the groups identified from
the O star sample in the same way.  The group diameters are defined as
$D=\frac{1}{2} (\Delta \alpha + \Delta \delta)$ following Battinelli
(1991), where $\Delta \alpha$ and $\Delta \delta$ represent the
maximum difference between members in RA and Dec, respectively, in
degrees of arc.  We also list the individual member stars for each
group in Tables~\ref{t_4} and \ref{t_5}, which are fully available in
the on-line edition.  The first two columns of Tables~\ref{t_4}
and \ref{t_5} give the RA and Dec of each star in decimal 
degrees (J2000.0), columns 3 -- 10 give the $UBVR$ magnitudes and
uncertainties from Massey (2002), and the last two columns give
the star ID from Massey (2002) and our newly-determined OB group ID.
Table~\ref{t_5} also lists the {\sl UIT} B5 magnitudes from
Parker {\etal}(2004) in columns 11 and 12.
Note that these group identifications represent
the results from two separate runs of the group-finding algorithm;
thus, if a star belongs to both samples, it may belong to groups in
both Table~\ref{t_4} and Table~\ref{t_5}.

\begin{figure*}
\epsscale{2.0}
%\plotone{plotctrfiga.ps}
\plotone{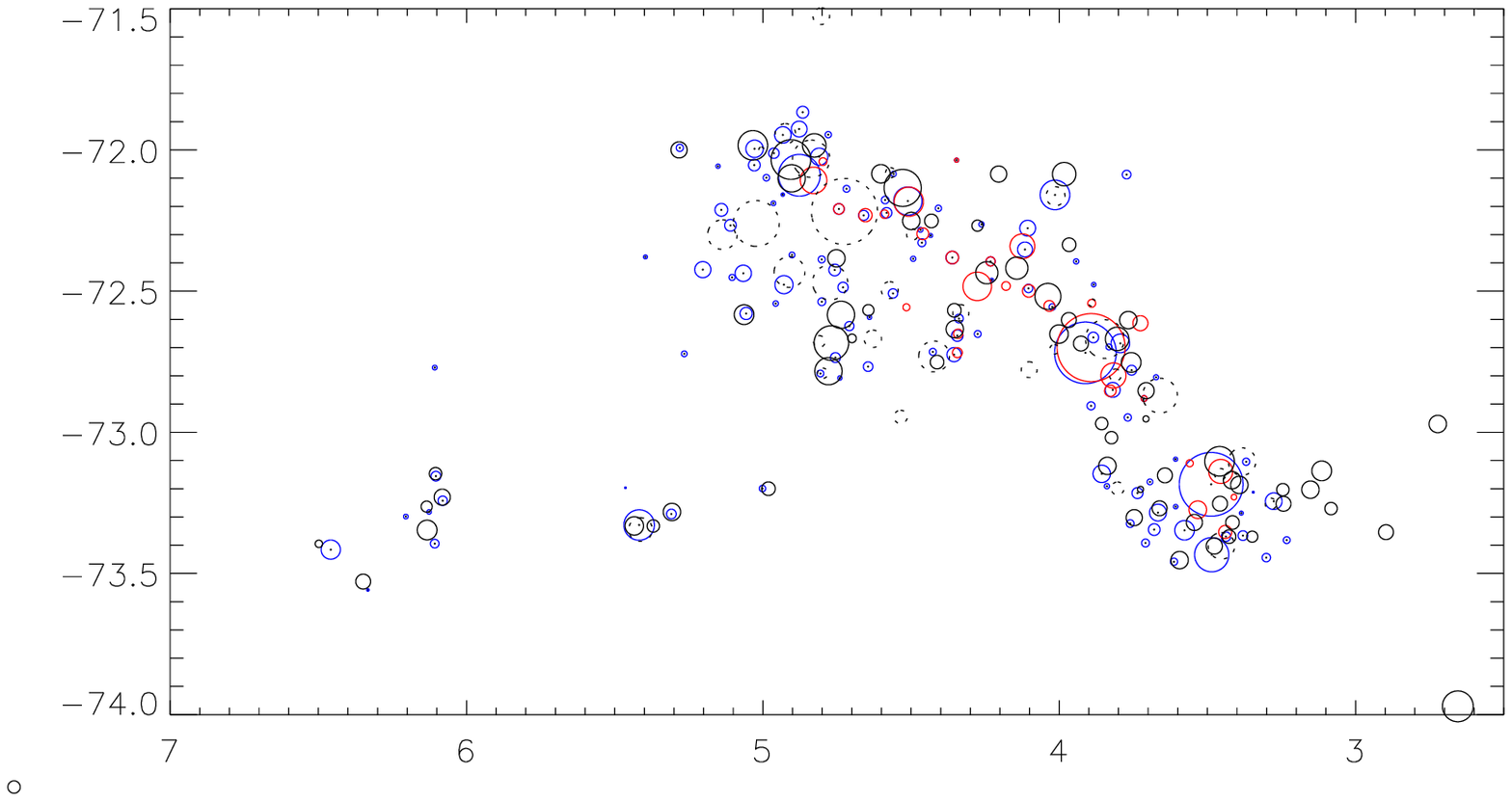}
\vspace*{-4in}
%\vspace*{-3.5in}
\caption{Centroids and sizes of associations identified by Hodge
(1985; solid black), Battinelli (1991; dotted black), our OB sample
(blue) and our O-star sample (red).  Note that all blue circles have a
dot at the center, thus overplotted identical circles can be
identified by small dots in the center of red circles.  (RA and Dec
given in degrees of arc.) 
\label{f_grouppos} }
\end{figure*}

\begin{figure*}
\vspace*{-1.0in}
\epsscale{1.0}
%\epsscale{0.5}
%\plotone{sizes_per_group.OB28.ps}
%\plotone{sizes_per_group.O34.ps}
\plotone{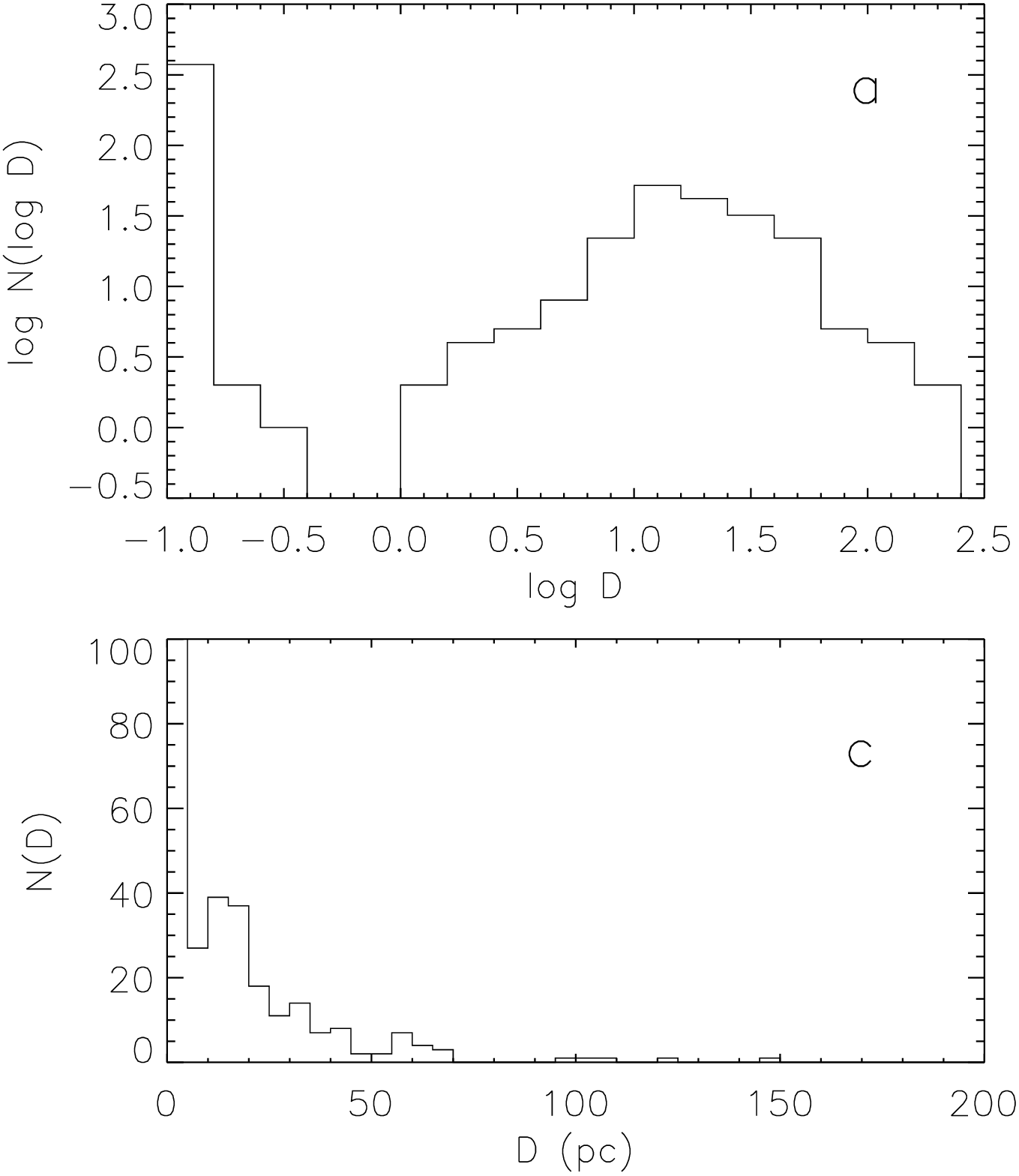}
\plotone{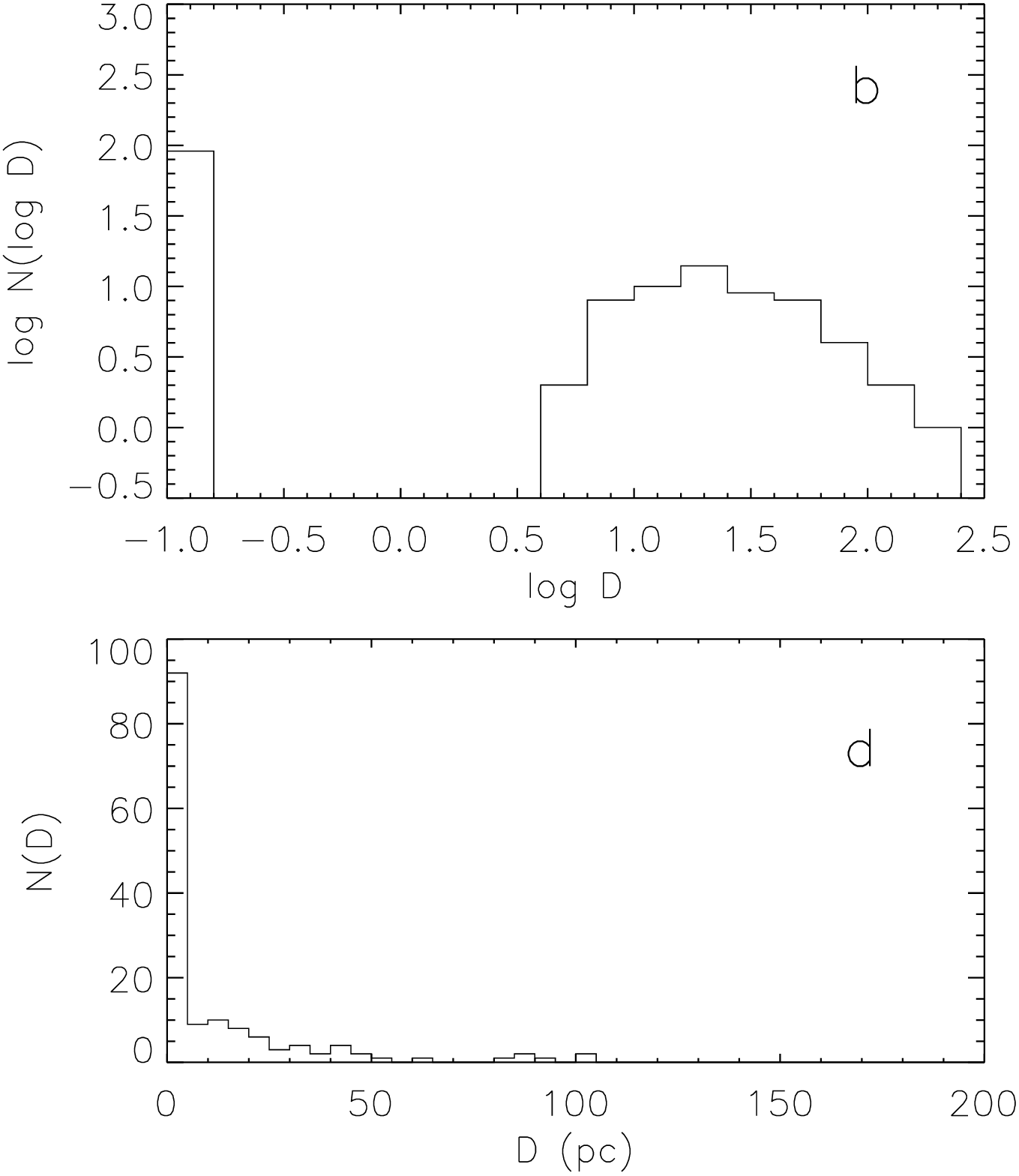}
%\vspace*{-0.5in}
\caption{Logarithmic size distribution for groups identified from the
candidate OB and O-star samples are shown in the top row (panels $a$
and $b$, respectively).  The linear distributions for the OB and
and O-star samples are shown in the bottom row (panels $c$ and $d$,
respectively).  The first bin in Figure~\ref{f_sizedistr}$c$ has a
maximum of 391.
\label{f_sizedistr} }
\end{figure*}

Figure~\ref{f_grouppos} shows the location of the group centroids
compared to those of associations identified by Hodge (1985; solid
black circles) and Battinelli (1991; dashed black circles).  Our OB
and O-star groups are indicated by the solid blue and red circles,
respectively.  The circle sizes correspond to the mean diameters $D$
from Tables~\ref{t_OBgrpctrs} and \ref{t_Ogrpctrs}.  For the Hodge
objects, the diameters are taken to be the mean of the dimensions in
RA and Dec given by Hodge (1985).  There is good general
correspondence between the positions of our groups and these earlier
catalogs; certainly at least as good as the correspondence between the
Hodge (1985) and Battinelli (1991) identifications.  It is apparent
that the net tendency from our smaller $d_s$ is for our associations
to be smaller than those in the earlier catalogs, and to break up some
of those associations into smaller groups.
Figures~\ref{f_sizedistr}$a,c$ and \ref{f_sizedistr}$b,d$ show the
size distributions for, respectively, the OB and O-star samples.  It
is apparent that, aside from individual stars, the peak in the
distribution is around 10--15 pc, compared to 50 pc for both the
Battinelli (1991) and Hodge (1985) catalogs.  Whereas Hodge (1985)
reports that the mean diameter of his sample is similar to the mean
for the LMC sample, our smaller characteristic value is consistent
with the lower luminosities of the SMC \hii\ regions and star-forming
regions compared to those of the LMC (e.g., Kennicutt, Edgar, \& Hodge
1989). 

\section{Field vs Clusters}

\begin{figure*}
\vspace*{-1.0in}
\epsscale{1.0}
%\epsscale{0.5}
%\plotone{number_per_group.OB28.ps}
%\plotone{number_per_group.O34.ps}
\plotone{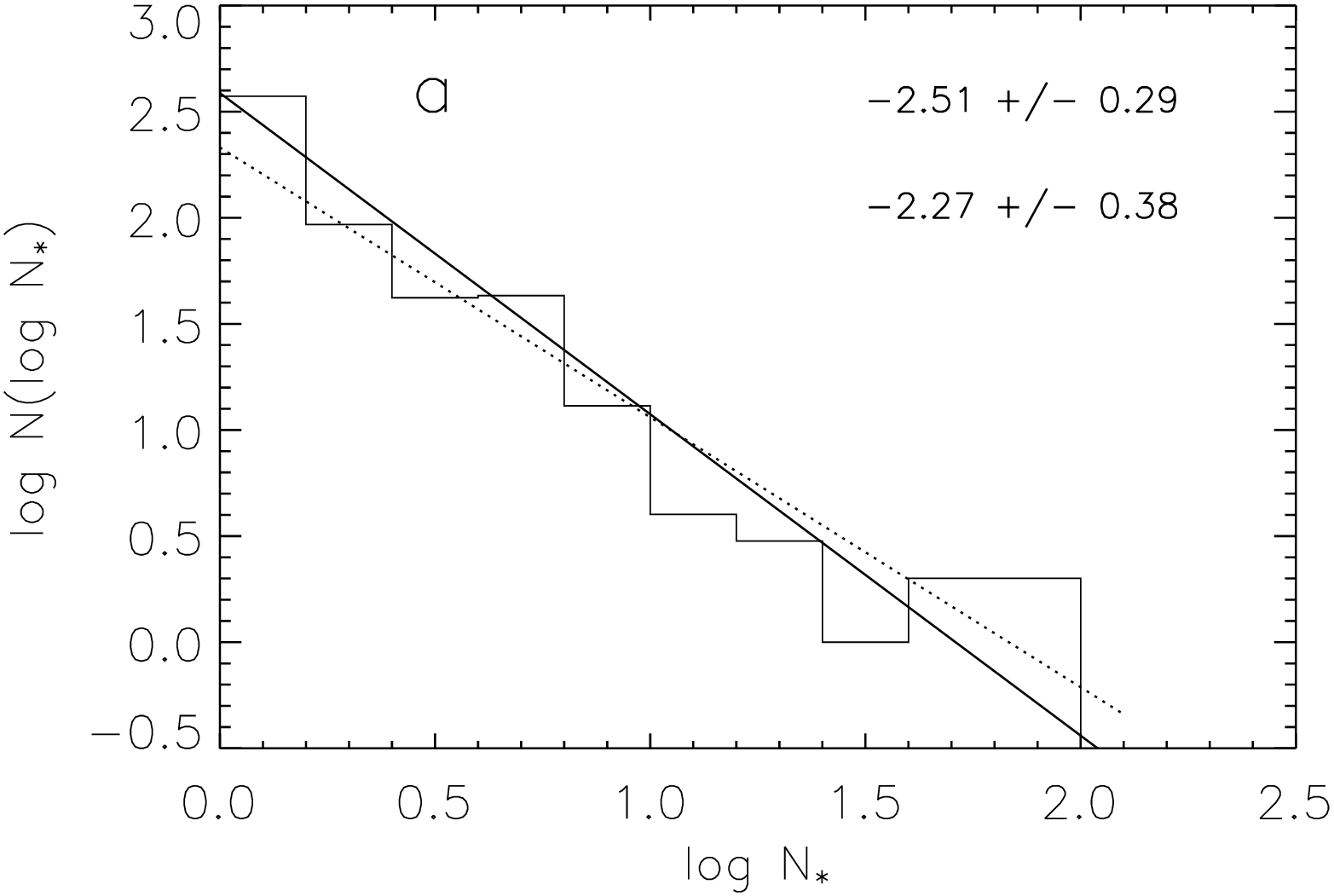}
\plotone{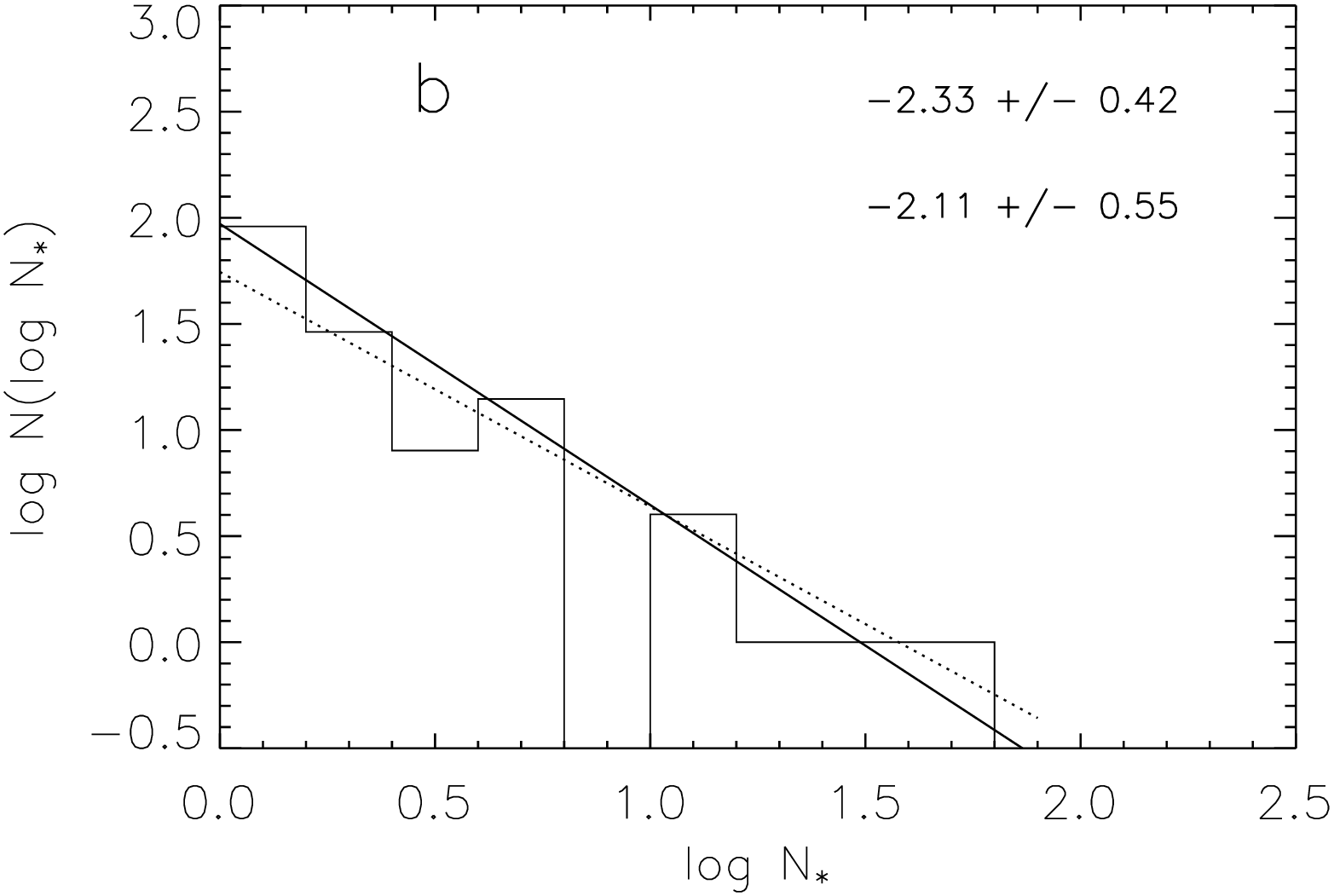}
\vspace*{-1.5in}
\caption{The distribution in \nstar\ for groups defined from the candidate OB
sample (panel $a$) and O-star sample (panel $b$).  Fits to the power-law
exponent ($-\beta$) are shown for the entire distributions (upper values)
and omitting the first bin of single stars (lower values).  Note that
the slope fitted in logarithmic space is $-\beta + 1$.
\label{f_nstar} }
\end{figure*}

%\twocolumn
Figure~\ref{f_nstar} shows the distribution in \nstar\ for the OB
groups (panel $a$) and O-star groups (panel $b$).  Two slopes are
shown fitted to the power-law distribution, weighted by the square
root of the bin value, $\sqrt{N(\log N_*)}$:  the solid line shows the
fit for the entire distribution, yielding $-2.51\pm 0.29$ and
$-2.33\pm 0.42$ for the OB and O-star samples, respectively; and the
dotted line shows the fit omitting the first bin, which corresponds to
single stars, resulting in fits of $-2.27\pm 0.38$ and $-2.11\pm 0.55$
for the OB and O-star samples, respectively.  The fits omitting the
single stars agree with the power-law slope of --2
(equation~\ref{eq_n*}), found for the \nstar\ distribution in a
variety of systems, as discussed in \S 1.  We again emphasize, as
mentioned above, that the fitted slope has a dependence on the
clustering distance $d_s$:  for large $d_s$, more stars are drawn into
the associations, causing the resultant slope of the clustering law to
be flatter relative to clustering defined by a small $d_s$.  Since we
seek a characteristic value for the slope of $N(N_*)$, it is important
to ensure that $d_s$ is in turn characteristic of the sample.  For the
$d_s$ in the extreme range of 20 -- 40 pc (see Figure~\ref{f_ds}$b$),
the O-star sample shows a variation in fitted slope, omitting the 
field stars, of $-2.79\pm 0.84$ to $-1.85\pm0.52$, respectively.

\begin{figure*}
%\vspace*{-0.5in}
\epsscale{2.0}
%\epsscale{1.0}
%\plotone{Oselect.ps}
\plotone{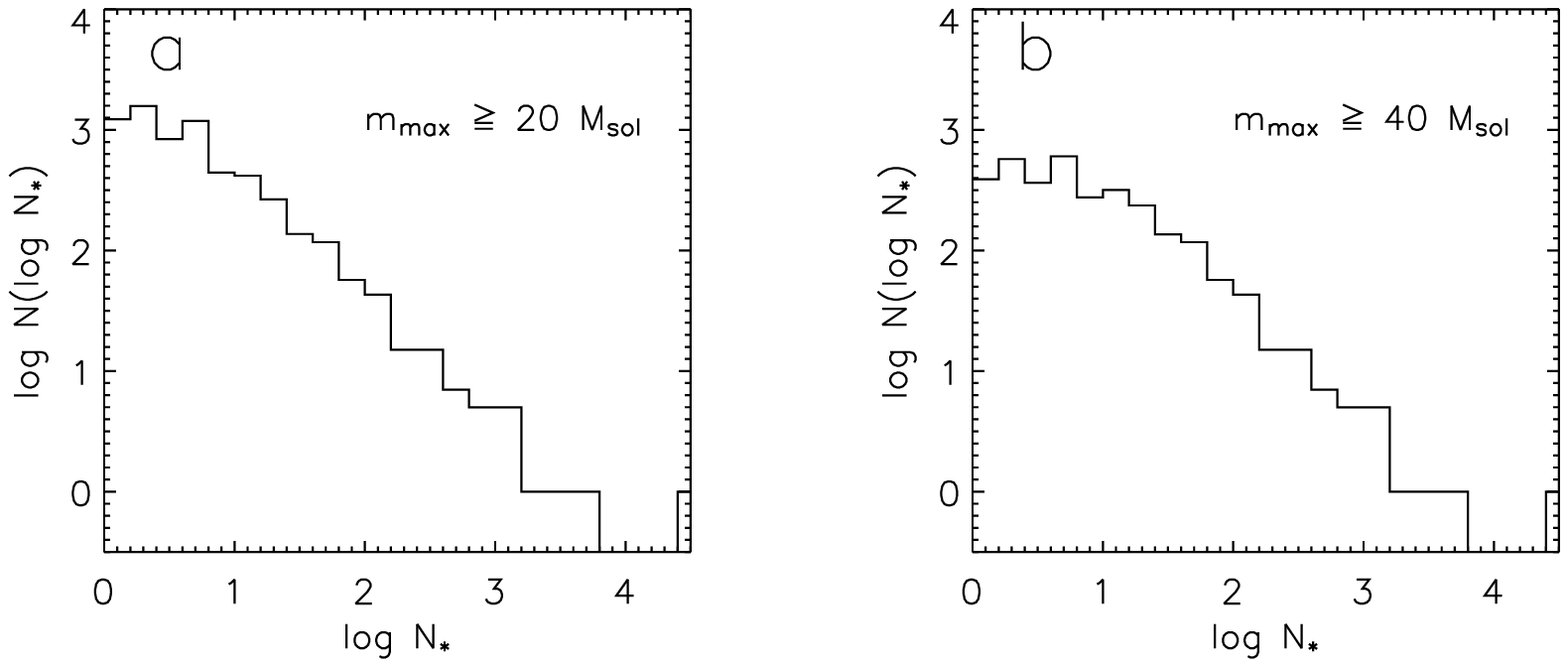}
\vspace*{-6.0in}
%\vspace*{-5.7in}
\caption{Monte Carlo models for distributions of $\log N_*$, selected
by the criterion that the maximum stellar mass in each cluster be at
least $m_{\rm max}\geq m_{\rm cut}$, showing $m_{\rm cut}=20\ \msol$ and
$40\ \msol$ in panels $a$ and $b$, respectively.
\label{f_cuteff} }
\end{figure*}

It is apparent that the single stars are slightly greater in number
than the power law for the remainder of the \nstar\ distribution,
particularly in the OB sample.  How significant is this excess in the
isolated, ``field'' stars?  We note that the magnitude of the excess
may be a lower limit, since the slope of the \nstar\ distribution
is expected to flatten slightly at the smallest values of \nstar, owing to
statistical effects, as follows.  We assume, as suggested above, that
the OB stars counted in the value of \nstar\ represent only the most
massive stars for a population of star clusters that are in reality
populated by a mass distribution described by a conventional IMF.
Since we identified only OB stars, our clusters are selected with the
criterion that each cluster contains a star of at least mass $m_{\rm cut}$.
The smallest, ``unsaturated'' clusters have a lower probability of
having their maximum stellar mass $m_{\rm max} \geq m_{\rm cut}$, so
they will be progressively missing from a sample of clusters selected
in this way (see Oey \& Clarke 2003 for a detailed discussion of
this effect).  

Figure~\ref{f_cuteff} shows Monte Carlo models that
demonstrate this effect.  \nstar\ is drawn from a power-law
distribution of slope --2 (equation~\ref{eq_n*}), with the individual
stellar masses drawn from a Salpeter (1955) IMF having $10 \leq m\leq
100\ \msol$.  Figure~\ref{f_cuteff}$a$ shows the distribution in
\nstar\ for 10,000 clusters having at least one star with mass $m_{\rm
max} \geq 20\msol$, and Figure~\ref{f_cuteff}$b$ is the same, but
selected for $m_{\rm max} \geq 40\msol$.  We see a flattening in the
distributions for small \nstar, and the flattening is stronger in
Figure~\ref{f_cuteff}$b$, since the probability of drawing a $40
\msol$ star is less than for a $20 \msol$ star. 

\begin{figure*}
\vspace*{-0.2in}
%\vspace*{-0.5in}
\epsscale{1.5}
%\epsscale{1.0}
%%\plotone{singlebkg.ps}
\plotone{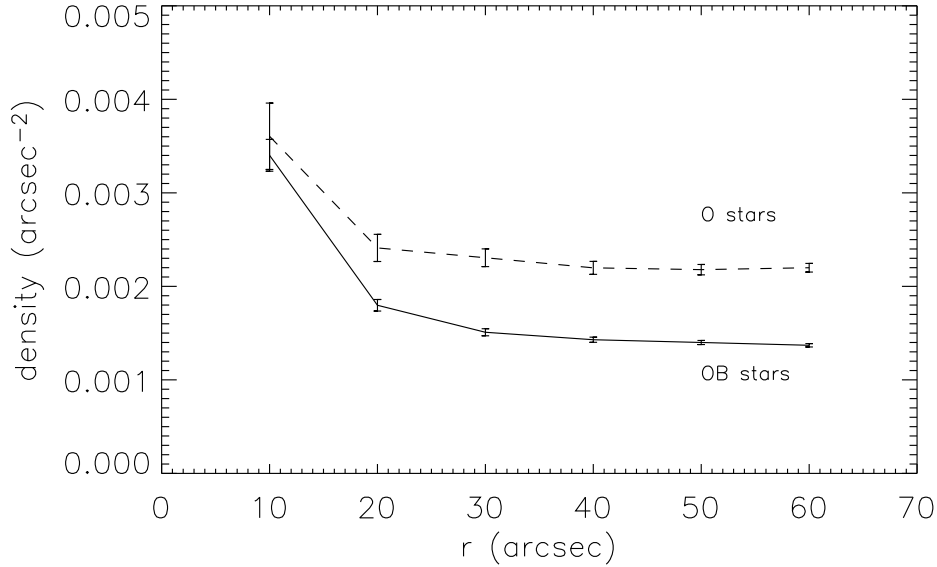}
\vspace*{-3.0in}
%\vspace*{-4.5in}
\caption{Mean stellar density (arcsec$^{-2}$) within radius $r$
(arcsec) from isolated massive stars in the candidate OB (solid line)
and O-star (dashed line) samples ($10\arcsec = 2.9$ pc).  The O-star
sample shows a higher background density since that sample is limited
to the SMC bar region (Figure~\ref{f_areas}).
\label{f_singlebkg} }
\end{figure*}

Figure~\ref{f_singlebkg} shows the stellar density of all catalogued
stars in the $UBVR$ survey as a function of radius around the
isolated, field massive stars found in the OB (solid line) and O-star
(dashed line) samples.  The errors are computed as $\sqrt{n}/\pi r^2$
for the total $n$ stars within radius $r$, omitting the central field
massive stars.  As seen in Figure~\ref{f_singlebkg}, 
it is apparent that the stellar density increases at small $r$.  This
is consistent with our suggestion that most massive field stars
represent the most massive component of groups of smaller stars, as is
expected from the stellar IMF.   

\begin{figure*}
%\vspace*{-0.5in}
\epsscale{2.0}
%\epsscale{1.0}
%%\plotone{unsatN.ps}
\plotone{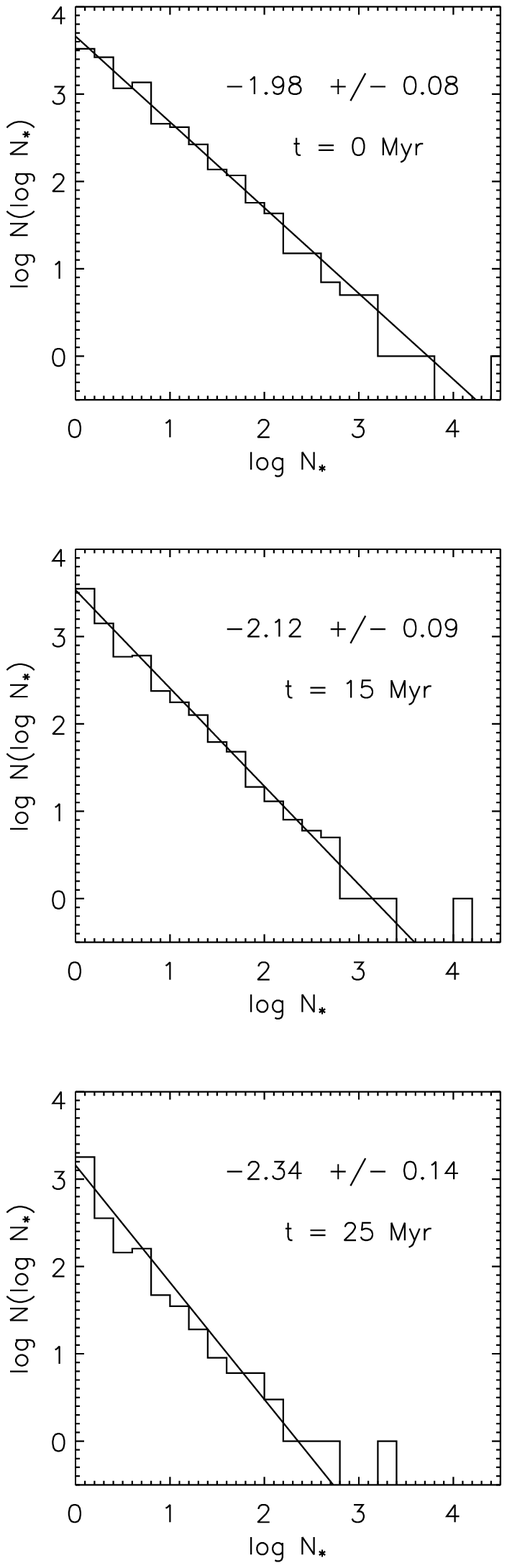}
%\vspace*{-2.0in}
% \vspace*{-0.3in}
\caption{Monte Carlo models for evolution in the $\log N_*$
distribution, assuming that all objects were created simultaneously.
Models for ages of 0, 15, and 25 Myr are shown, with fitted power-law
slopes.  As in Figure~\ref{f_nstar}, the linear power-law exponents
are shown, while the fitted slope is $-\beta+1$ in logarithmic space.
\label{f_burst} }
\end{figure*}

Another effect that can offset the statistical flattening in the form
of the \nstar\ distribution for small \nstar, is evolution.  In the most
extreme situation, we consider that all the OB associations in
the SMC were formed together in a single burst of global star
formation.  Staveley-Smith et al. (1997) find that the \hi\ shells
identified in the SMC appear to have a narrow age distribution around
5~Myr.  We described above that the unsaturated, lowest-\nstar\
clusters have a lower average stellar mass (e.g., Oey \& Clarke 2003).
Since stellar lifetimes are longer for lower-mass stars, we therefore
expect that these unsaturated clusters will tend to last longer, on
average, than the statistically fully-sampled, or saturated, objects.
Thus, we may expect that the distribution in \nstar\ steepens with
time.  

We constructed Monte Carlo simulations of such an aging burst of clusters.
Figure~\ref{f_burst} shows the models for a population of clusters at
0, 15, and 25 Myr after their simultaneous formation.  We used the
same IMF parameters are before, and the stellar ages are from the grid
of Charbonnel et al. (1993) for SMC metallicity.  Each of the model
\nstar\ distributions is fitted with a power-law slope, shown in
Figure~\ref{f_burst}.  We see that, while it is difficult to discern the
slope steepening over the entire sample after 15 Myr, and even at 25
Myr, it is apparent that the aging effect is most pronounced in the
smallest $N_*$ bins, and that the \nstar=$1$ bin becomes the most
disproportionately enhanced.  For continuous creation of the clusters,
the net observed slope will be intermediate between the forms in
Figure~\ref{f_cuteff} and Figure~\ref{f_burst}.  

The modest observed excess of single stars is also likely to be caused
in part by a contribution to their population by runaway OB stars.
If runaway OB stars originate in associations, however, then this
contribution must be small, since runaways correspond to about 3\% and
20\% of field early B and early O stars, respectively (Blaauw 1961).
Runaways will contribute primarily to the single-star population,
while binary runaways represent $\lesssim 19$\% of all runaways (Gies
1987).  Binary runaways are likely to be tight pairs that remain
unresolved in the 0.67 pc px$^{-1}$ survey resolution.
% Figure~\ref{f_nstar}$b$ is consistent with a modest enhancement in
% both the \nstar$=1$ and \nstar$=2$ bins.

It is also possible that spurious selection of candidate stars, owing
to the coarse photometric criteria, are a factor.  We showed in \S 2
that there may be a net of about 15\% of the O-star sample that consist
of spuriously selected B stars; while in principle, these should be
distributed proportionately between clusters and the field, it is
possible that crowding effects in clusters favor field selection.
If so, owing to the longer lifetimes of B stars, they could contribute
disproportionately to the field star population.  However, we
emphasize that this requires a significant variation in the spatial
distribution of only the spurious candidates, between field and clusters.

\begin{figure*}
\epsscale{2.0}
%\epsscale{1.0}
%%\plotone{plotctrfigb.ps}
\plotone{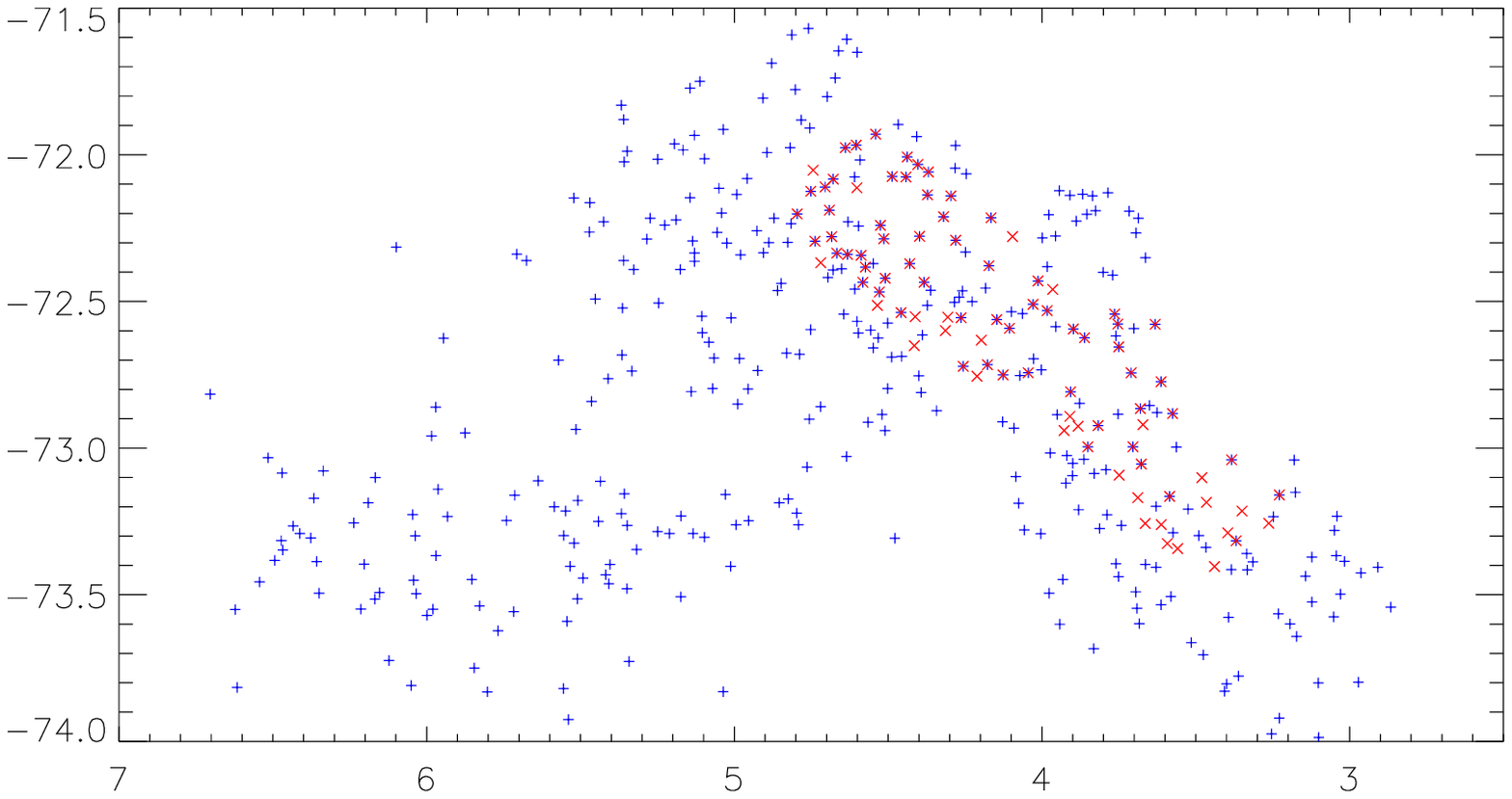}
\vspace*{-4in}
\caption{Spatial distribution of isolated field stars for the OB
sample (plus symbols), and O-star sample (cross symbols).  Scale and
axes are as in Figure~\ref{f_grouppos}.
\label{f_fieldpos} }
\end{figure*}

The most important factor in enhancing the field star population,
however, is probably
the strong variation in star formation density across the SMC.  The
highest star formation density occurs in the SMC bar, where, as seen
in Figure~\ref{f_grouppos}, most of the associations are located.
Individual massive field stars, on the other hand, have a much more
uniform distribution, as seen in Figure~\ref{f_fieldpos}.  The uneven
star formation density distribution in this galaxy therefore favors
the field stars and enhances the \nstar$=1$ bin for the OB sample in
Figure~\ref{f_nstar}.  The effect is much less pronounced in the
O-star sample, which is limited to the SMC bar region.

Considering these effects, it is apparent that there is no strong
variation or change in character of the power-law distribution seen in
Figure~\ref{f_nstar} for the smallest values of \nstar.  The empirical
samples (Figure~\ref{f_nstar}) are much smaller than those in the
models and therefore have significantly poorer statistics in $\log
N(\log N_*)$.  Thus, the modeled effects will be more difficult to
discern in the data.  Since our results are largely consistent with a
single intrinsic power-law form for the clustering law, this
suggests that the massive field star population simply represents an
extension of the massive cluster population extending down to $N_*=1$.
There is no evidence that the majority of massive field stars
originate from a mode of star formation that is different from those
in associations.  However, further studies of massive field and
cluster populations in other environments are necessary to confirm
the generality of these results.

\section{Discussion}

For a universal IMF, the constant slope of the \nstar\ distribution
extending to the field stars has profound consequences for their
global feedback influence in galaxies.  We can now quantify the 
assertion that most massive stars form in associations:  for an
\nstar\ distribution given by equation~\ref{eq_n*}, the 
total number of OB stars is, 
\begin{equation}\label{eq_n*sum}
N_{*,\rm tot} \propto \sum_{N_*=1}^{N_{*,\rm up}} N_* \cdot N_*^{-2} \quad ,
\end{equation}
where  $N_{*,\rm up}$ refers to the upper-limit, maximum cluster
of stars.  Equation~\ref{eq_n*sum} corresponds to
a divergent harmonic series which can be approximated for
large $N_{*,\rm up}$ as:
\begin{equation}\label{eq_n*sumsoln}
N_{*,\rm tot} \simpropto \  \ln N_{*,\rm up} + \gamma \quad ,
\end{equation}
where $\gamma \simeq 0.5772$ is the Euler-Mascheroni constant.  Thus
the fraction of \nstar$=1$ field stars is $(\ln N_{*,\rm up} +
\gamma)^{-1}$ of the total $N_{*,\rm tot}$.  For our OB and O star
samples, respectively, $\log N_{*,\rm up} \simeq 2.0$ and 1.8,
yielding 19\% and 21\% fractions for the field stars.  

Counting the actual stars in our OB and O star samples, we find that
374 and 91 candidates, respectively, had no massive companions within
the clustering radius.  These correspond to 28\% and 24\%, reflecting
the excess found above.  These fractions are about a factor two lower 
than the finding by Parker et al. (2001) that over half of their {\it
UIT}-selected candidate O stars are outside catalogued association
boundaries in the Large Magellanic Cloud (LMC).  The results for both
the SMC and LMC may be odds with the results of Massey et al. (2002),
who found a much steeper IMF slope for field vs cluster massive stars:
a steeper IMF slope in the field would be manifested as a smaller
number of field OB stars, yet the slope of the \nstar\ distribution
tends to be steeper than expected, rather than flatter.  If the field
star IMF is indeed steeper than in clusters, then the clustering law
must also steepen substantially for the smallest clusters, in such a
way as to compensate for a steep IMF in our data.  Further
investigation is necessary to resolve this issue. 

In the meantime, the data appear to be broadly consistent with the
simpler scenario of, {\it simultaneously}, a universal IMF and universal
clustering law given by equation~\ref{eq_n*}.  As emphasized by
McKee \& Williams (1997), the total OB star population,
and thus the fraction of isolated field massive stars, is driven by
$N_{*,\rm up}$ with an inverse logarithmic dependence.
For maximum $N_{*,\rm up}$ ranging between 10 and $10^6$,
equation~\ref{eq_n*sumsoln} yields a fraction of field OB
stars ranging from 35\% to 7\%, respectively.

The clustering law in equation~\ref{eq_n*} has important consequences
for feedback, implying that the interstellar porosity caused by the
formation of superbubbles and supernova remnants has equal relative
contributions from objects of all sizes 
(Oey \& Clarke 1997).  And, since the Str\"omgren volume $V_s\propto N_*$,
it also implies that a strong majority of the field massive stars likely
contribute to ionizing the diffuse, warm ionized medium, which
constitutes about 40\% of the total \Ha\ luminosity in star-forming
galaxies (e.g., Walterbos 1998).  Our result is quantitatively
consistent with the result of Hoopes \& Walterbos (2000) that field OB
stars can power 40\% $\pm$ 12\% of the warm ionized medium in M33,
where the fraction of field OB stars is likely around 15\%.

\section{Conclusion}

We find no evidence that the field massive stars in
the SMC are formed by a fundamentally different star-forming process.
Rather, we find that the continuous power-law distribution in \nstar\
down to \nstar$=1$ strongly suggests that the star-forming process is
continuous from rich clusters to poor groups, apparently for all
ensembles that form OB stars.  The data are consistent with the model
that single, field OB stars are usually the most massive member of a
group of smaller stars, as expected from the universal \nstar\
distribution (equation~\ref{eq_n*}).  These results are consistent with
the {\it simultaneous} existence of a universal IMF and universal
$N_*^{-2}$ clustering law.  These joint universal power laws
imply that field OB stars constitute roughly 35\% to 7\% of the total
massive star population, with an inverse logarithmic dependence on
$N_{*,\rm up}$ of the the most populous cluster.  Thus, the fraction is
dependent on galaxy size and/or star formation rate.  The contribution
of these field stars to the ionized volume in the warm ionized
medium is likely to scale according to their relative fraction.  The
universal clustering law also implies equal relative contributions by
superbubbles of all sizes to the interstellar porosity.

\acknowledgments

We are pleased to acknowledge discussions with Ren\'e Walterbos,
Cristiano Porciani, Todd Small, and especially, Chris McKee and the
anonymous referee.  We also thank Phil Massey for generously providing
the $UBVR$ data in advance of publication.  This work was supported by
the NASA Astrophysics Data Program, grant NAG5-10768. 

% \vfill\eject

% |

% \onecolumn

\end{document}